\documentclass[preprint,amsmath,amssymb,aps,showkeys,showpacs,pra]{revtex4-1}

\usepackage[english]{babel}
\usepackage{graphicx}
\usepackage{graphics}
\usepackage{amsmath}
\usepackage{dcolumn}
\usepackage{amssymb}
\usepackage{bm}
\usepackage[latin1]{inputenc}
\usepackage{graphicx,color}

\begin{document}
\title{Weyl fermions in a family of G\"{o}del-type geometries  with a  topological defect}

\author{G. Q. Garcia}
\email{gqgarcia@fisica.ufpb.br}
\affiliation{Departamento de Física, Universidade Federal da Paraíba, Caixa Postal 5008, 58051-970, João Pessoa, PB, Brazil.}

\author{J. R. de S. Oliveira}
\email{jardson.ricardo@gmail.com}
\affiliation{Departamento de Física, Universidade Federal da Paraíba, Caixa Postal 5008, 58051-970, João Pessoa, PB, Brazil.}

\author{C. Furtado}
\email{furtado@fisica.ufpb.br} 
\affiliation{Departamento de Física, Universidade Federal da Paraíba, Caixa Postal 5008, 58051-970, João Pessoa, PB, Brazil.} 


\begin{abstract}
In this paper we study Weyl fermions in a family of G\"odel-type geometries  in Einstein general relativity.  We also consider that these solutions  are embedded in a topological  defect background. We solve the Weyl equation and find  the energy eigenvalues and eigenspinors for all three cases  of  G\"odel-type geometries  where a topological defect is passing through them. We show that the presence of  a topological in these geometries contributes to modification of the spectrum of energy. The energy zero  modes for all three  cases of the G\"odel geometries are discussed.
\end{abstract}


\maketitle

\section{Introduction}

The G\"odel \cite{godel} solution of the Einstein  equation is the first example of cosmological universe  with rotating matter. It was obtained considering  a   cylindrically symmetric  stationary solution  of the equations of  the general relativity.   The solutions of this class  are characterized by the presence  of closed timelike curves (CTCs). Hawking \cite{hawking} investigated the presence of   CTCs  and  had   conjectured that the presence of CTCs  is physically inconsistent. Rebou\c cas {\it et al} \cite{reboucas,reboucas2,reboucas3}  investigated the G\"odel-type solution in general relativity and  Riemann-Cartan theory of gravity. They presented a detailed study of  the problem of causality in  this family of spacetimes, and then, established  the following set of  classes of solutions: (i) solutions where there are no CTCs; (ii) solutions where there  is a sequence of  alternating causal and non-causal regions; finally, the case (III) where a class of solutions  involves  only one non-causal region. In Ref. \cite{dabrowski}, Dabrowski   had investigated  the criteria  necessary for  the existence of CTCs, to obtain this criteria he have made use of two quantities,  called the   super-energy and super-momentum. Recently,  the appearance of CTCs in  the G\"odel-type  spacetime was investigated in  the framework of   string theory \cite{dan,bertolami} in Ref. \cite{barrow}. Others interesting properties of the G\"odel solution  were investigated  by  Barrow in the Ref. \cite{barrow2}. This family  of solutions is given by the line element:
\begin{eqnarray}
ds^2 = -\left(dt + \frac{H(r)}{D(r)} d\phi\right)^{2} + \frac{1}{D^{2}(r)}\left(dr^{2} + J^{2}(r) d\phi^{2}\right) + dz^{2}, 
\label{1.1}
\end{eqnarray}  
where for each case, we have different functions $H(r)$, $J(r)$ and $D(r)$: {\bf(i)} for the case of Som-Raychaudhuri spacetime we have $D(r) = 1$, $H(r)=\alpha\Omega r^{2}$ and $J(r) = \alpha r$. {\bf(ii)} we have $D(r) = \left(1 +\frac{r^{2}}{4R^{2}}\right)$, $H(r) = \alpha\Omega r^{2}$ and $J(r) = \alpha r$, for the case of G\"odel- type spacetime with spherical symmetry. {\bf(iii)}  finally, for hyperbolic G\"odel- type spacetime we  have $D(r) = \left(1 - l^2 r^2\right)$, $H(r) = \alpha\Omega r^{2}$ and $J(r) = \alpha r$.   The system of coordinates  $(t,\,r,\,\phi,\,z)$ are defined in the ranges: $0\leq r <\infty$, $0\leq\phi\leq 2\pi$ and $-\infty<(z,t)<\infty$. The parameters $\Omega$ and $\alpha$ are related with the vorticity and angular deficit  arising from  the topological defect, respectively, since it is associated with the deficit of angle $\alpha=(1-4\Theta)$, with $\Theta$  being the mass per unit length of the cosmic string, and it assumes values in the range $0<\alpha<1$.  In geometric theory of defect in condensed matted  for a negative disclination, which are characterized by angular excess where the parameter $\alpha$ can assumes values $\alpha>1$. Note if you make the following change of variable $\theta=\alpha\phi$ the theta are defined in the range $0<\theta<2\pi\alpha$. the introduction of defects have topological nature and it presence in this geometry do not removed by simple coordinate transformation. 
Several studies of the physical  problems involving G\"odel-type spacetimes have been developed in recent years.    These geometries   have been studied from the  point of view of the equivalence problem techniques in  the Riemannian G\"odel-type spacetimes \cite{reboucas}  and for Riemann-Cartan G\"odel-type spacetimes \cite{joeljmp,joel,joelrebou}.  Recenty, a large number of issues related to rotating G\"odel solutions in general relativity as well as  in alternative  theories of gravitation have been studied,  for example:  the hybrid metric-Palatini gravity \cite{reboupala}, the Chern-Simons modified gravity theory \cite{furgodel,furgodel3}, the Horava-Lifshitz theory of gravity \cite{furgodel2,joelhorava} and the Brans-Dick theory of gravity \cite{porfi}. In a recent paper \cite{everton}, the electronic properties of spherical symetric carbon molecule -- {\it Buckminsterfullerene} --  were analyzed employing a geometric model based  on  the spherical     G\"odel spacetime. Figueiredo {\it et al} \cite{figueiredo}  investigated  the scalar and spin-$1/2$  particles in G\"odel  spacetimes  with positive, negative  and   zero curvatures. The relationship between the  Klein-Gordon  solution in a class of G\"odel solutions in general relativity with Landau  levels in a curved spaces \cite{comtet,dunne}  was investigated by  Drukker {\it et al} \cite{fiol,bfiol}. This analogy was also observed by Das and  Gegenberg \cite{gegenberg}  within studying the quantum dynamics of scalar particles in the Som-Raychaudhuri spacetime (G\"odel flat solution) and compared with the Landau levels in the flat space.  The quantum dynamics  of a scalar quantum particle in  a class of the G\"odel solutions with a presence of a cosmic string  and  the solutions of  Klein-Gordon  equation in the Som-Raychaudhuri spacetime have also been studied  in Refs. \cite{josevi,epjpchina}.  In Ref. \cite{vilalba}   Villalba  had obtained solutions of Weyl equation in a non-stationary G\"odel-type cosmological universe.  The Weyl equation was  studied for a family of metrics of the G\"odel-type by Pimentel and collaborators \cite{pimentel}. They  have solved the Weyl equation for a specific case of the  G\"odel solution. In  the  recent article  \cite{sandro} Fernandes {\it et al.} have solved the Klein-Gordon equation for a particle confined in two concentric thin shells in G\"odel, Kerr-Newman and FRW spacetimes with the presence of  a topological defect passing through them. Havare and  Yetkin \cite{havare} have studied the solutions of photon equation in stationary G\"odel-type and G\"odel spacetimes. Several  equations for different spins have been studied in the G\"odel universe \cite{cohem,hisco,mash,pimentel1}. 

 Continuing these studies, in this article we investigate the Weyl equation in  a family of G\"odel-type metrics in the presence of a topological defect.    
We obtain the solutions of  the Weyl  equation in this set of G\"odel geometries. Recently, a series of studies have been made with the purpose of investigating the quantum and classical dynamics of particles in curved spaces with topological defects and  the possible detection of this defect.  In Ref. \cite{gabriel} the quantum dynamics of Dirac fermions in G\"odel-type solution in  gravity with torsion was investigated, and we have observed that  the presence of torsion in the space-time yields new contributions to the relativistic spectrum of energies of massive Dirac fermion, and that the presence of the topological defect modifies the degeneracy  of energy levels.  Another conclusion is that the  torsion  effect on the allowed energies corresponds to the splitting of each energy level in a doublet.    The purpose of this contribution is to investigate the influence of curvature, rotation  of G\"odel-type metrics and  topology introduced by the topological defect in the  quantum dynamics of Weyl fermions.  We investigated influence of  Gödel-type geometries   of positive, negative and flat curvature that contain a  topological defect in the eigenvalues and eigenfunctions of Weyl fermions. The possible application in Condensed matter systems are discussed. In this system the quasiparticle has behaviour similar the Weyl fermions an it travel by the "speed of light". Ins this system the model are described by Weyl fermion where Fermi velocity play de role of the light velocity in this effective theory. Recently in condensed matter physics there are systems that quasi-particles behave with a massless fermions or  Weyl fermions. Systems such as graphene, Weyl semi-metals and topological insulator. In this way, the study carried out in this article can be used to investigate the influence of rotation, curvature and topology in the condensed matter systems described by massless fermions.  The results obtained for the present case of Weyl fermions in a family of G\"odel-type metrics in Einstein relativity theory  are  quite different  from the previous result obtained for Dirac fermions with torsion in \cite{gabriel}. We claim that the studies of these problems in the present paper can be  used to investigate the influence of the topological defect in the G\"odel-type  background metrics. The approach applied  in this paper  can be used to investigate the influence of disclinations (cosmic strings)  in condensed matter systems as well as to investigate  the Hall effect in spherical droplets \cite{halp} with rotation and with the presence of disclinations.  Recently, in ref.\cite{everton} one of us have used a similar approach to investigate the influence of rotation in fullerene molecule where the  in this model was the rotation is introduced via a three-dimensional Godel-type metric. In this way,  we can study using this approach proposed here in  condensed matter system described by curved geometries  with rotation. In this paper, we analyse  the relativistic quantum dynamics of a  massless fermion  in  the presence of a topological defect in a class of G\"odel-type metrics. We solve Weyl  equation in Som-Raychaudhury, spherical and hyperbolic background metrics  pierced by  a  topological defect. We found  the eigenvalues of energy in all three cases and observe  their similarity with Landau levels for a  massless spin-$1/2$ particle. We also observe that presence of the topological defect breaks the degeneracy of the relativistic energy levels, and the eigenfunctions depend on the parameter that characterizes the presence of the topological defect  in  these background metrics.   The possibility of zero mode for the eigenvalues of the Weyl spinor are discussed and the physical implications are analysed for for all case three class of geometries investigated in this paper.

This contribution is organized as follows: in section \ref{secgod}, we present the Weyl equation in a geometry of G\"odel-type metric  pierced by  a topological defect  In an Einstein theory of relativity. The Weyl  equation in the background   of the G\"odel-type metric is written for the flat space in the section \ref{subsec1},  for the spherical one in the section \ref{subsec2} and  for the hyperbolic one in the section \ref{subsec3}. The relativistic bound state solutions to the Weyl equation in G\"odel-type background metric   are investigated; finally, in section \ref{sec3}, we present the conclusions. In this paper, we shall use natural units $\left(\hbar=c=G=1\right)$.

\section{Weyl fermions in G\"odel-type solutions in General Relativity}\label{secgod}
$\,\,$

In this section we will introduce the Dirac equation in Weyl representation on a curved background. Following the theory of  spinors in curved spacetime \cite{naka, weinberg, cartan1},  to do it, one must extend the partial derivative $\partial_\mu$  up to a covariant derivative $\nabla_\mu = \partial_\mu + \Gamma_\mu(x)$.  So that, we can write the equations for massless spin-$\frac{1}{2}$ field in the follow way,   
\begin{eqnarray}
i\gamma^\mu\nabla_\mu\psi = 0;
\label{2.1}\\
\left(1 + \gamma^{5}\right)\psi = 0;
\label{2.1.b}
\end{eqnarray}
here we have that $\gamma^\mu = \gamma^a e_{a}^{\ \mu}(x)$ are the gamma matrices in Weyl representation and $\gamma^5 = i\gamma^{0}\gamma^{1}\gamma^{2}\gamma^{3}$. 

Let us consider fermions in a G\"{o}del-type spacetime  of a general form. The metric of this curved spacetime is given by equation (\ref{1.1}). If we want to study the relativistic dynamics of fermions in the Som-Raychaudhuri spacetime, we  must introduce spinors in a curved spacetime  \cite{naka, weinberg, cartan1}. For this purpose, we will define the spinors via a noncoordinate basis $\hat{\theta}^a = e^{a}_{\ \mu}(x) dx^\mu$, where the components of tetrad $e^{a}_{\ \mu}(x)$ must obey the following relation $g_{\mu\nu}=e^{a}_{\ \mu}(x)e^{b}_{\ \nu}(x)\eta_{ab}$. The tensor $\eta_{ab}=diag(-+++)$ is the Minkowski tensor. We still can define the inverse tetrad  from the relation $dx^\mu = e^{\mu}_{\ a}(x)\hat{\theta}^a$, so  that $e^{a}_{\ \mu}(x)e^{\mu}_{\ b}(x)=\delta^{a}_{\ b}$ and $e^{a}_{\ \mu}(x)e^{\nu}_{\ a}(x)=\delta^{\nu}_{\ \mu}$. Thus, for the metric (\ref{1.1}),  the tetrad and its inverse are defined as
\begin{eqnarray}
e^{a}_{\ \mu}(x) = \left(\begin{array}{cccc} 1 & 0 & \frac{H(r)}{D(r)} & 0\\ 0 & \frac{1}{D(r)} & 0 & 0\\ 0 & 0 & \frac{J(r)}{D(r)} & 0\\ 0 & 0 & 0 & 1\\ 
\end{array}\right);\ e^{\ \mu}_{a}(x) = \left(\begin{array}{cccc} 1 & 0 & -\frac{H(r)}{J(r)} r & 0\\ 0 & D(r) & 0 & 0\\ 0 & 0 & \frac{D(r)}{J(r)} & 0\\ 0 & 0 & 0 & 1\\ 
\end{array}\right).
\label{2.4}
\end{eqnarray}

From solutions of Maurer-Cartan structures $d\hat{\theta}^a + \omega^{a}_{\ b}\wedge\hat{\theta}^b = 0$ \cite{cartan2, carroll}, we can obtain the 1-forms connections $\omega^{a}_{\ b}(x) = \omega^{\ a}_{\mu \ b}(x)dx^\mu$, with $d$  is exterior derivative and the symbol $\wedge$ is the wedge product. The object $\omega^{\ a}_{\mu \ b}(x)$ is the spinorial connection. Thus, the  non-zero components of spinorial connections are:
\begin{eqnarray}
\omega^{\ 0}_{\phi \ 1}(x) &=& -\omega^{\ 1}_{\phi \ 0}(x) = \frac{D}{2}\frac{d}{dr}\left(\frac{H}{D}\right) ;\nonumber\\
\omega^{\ 0}_{r \ 2}(x) &=& -\omega^{\ 2}_{\phi \ 0}(x) = -\frac{D}{2J}\frac{d}{dr}\left(\frac{H}{D}\right);\nonumber\\
\label{2.5}\\
\omega^{\ 1}_{t \ 2}(x) &=& -\omega^{\ 2}_{t \ 1}(x) = -\frac{D^{2}}{2J}\frac{d}{dr}\left(\frac{H}{D}\right);\nonumber\\
\omega^{\ 1}_{\phi \ 2}(x) &=& -\omega^{\ 2}_{\phi \ 1}(x) = -\left[\frac{DH}{2J}\frac{d}{dr}\left(\frac{H}{D}\right) + D\frac{d}{dr}\left(\frac{J}{D}\right)\right].\nonumber
\end{eqnarray}

Using the equations (\ref{2.1}) and (\ref{2.1.b}), we can use the chiral representation \cite{pal}, so that the spinorial connection will be $\Gamma_\mu(x) = \frac{1}{8}\omega_{\mu ab}(x)\left[\sigma^a,\sigma^b\right]$, where $\sigma^a$ are the Pauli matrices, and $\omega_{\mu ab}(x)$ are the 1-form connections related with the curvature of manifold given by equations (\ref{2.5}). Thus, the covariant derivative will be 
\begin{eqnarray}
\nabla_\mu = \partial_\mu + \frac{1}{8}\omega_{\mu ab}(x)\left[\sigma^a,\sigma^b\right].
\label{2.2}
\end{eqnarray}

As a result, the  Weyl equation will assume the following form:
\begin{eqnarray}
i\sigma^a e^{\ \mu}_{a}\left(\partial_\mu + \Gamma_\mu\right)\psi = 0 
\label{2.6}
\end{eqnarray} 

Then, using  the equations (\ref{2.5}) and (\ref{2.6}), we obtain the Weyl  equation in G\"odel-type spacetime in the following form:
\begin{eqnarray}
i\sigma^0\frac{\partial\psi}{\partial t} &+& i\sigma^1\left(D(r)\frac{\partial}{\partial r} -\frac{D(r)}{J(r)}\frac{\omega^{\ 1}_{\phi\ 2}}{2} +\frac{H(r)}{J(r)}\frac{\omega^{\ 1}_{t\ 2}}{2}\right)\psi +\nonumber\\
 &+& i\sigma^2\left(\frac{D(r)}{J(r)}\frac{\partial}{\partial \phi} -\frac{H(r)}{J(r)}\frac{\partial}{\partial t}\right)\psi + i\sigma^3\left(\frac{\partial}{\partial z} - i\nu A_{z}\right)\psi = 0.
\label{2.7}
\end{eqnarray}
We solve this equation for all three  possibilities for  functions $H(r)$, $J(r)$ and $D(r)$ for  G\"odel-type spacetimes.
After  we have described the Weyl equation in a G\"{o}del-type spacetime context, we are looking for the energy levels of  a massless  fermion in a G\"{o}del-type spacetime. We will solve  the Weyl equation for three classes of G\"{o}del-type metric: in the Som-Raychaudhuri spacetime, the G\"{o}del-type spacetime with spherical symmetry and the hyperbolic G\"{o}del-type spacetime.  

Note that the axial vector coupled with the  $z$-component of the momentum in equation (\ref{2.7}), for massless fermions, is a gauge field,  with $A_{z} = \Omega$ is a constant field, and $\nu = \frac{1}{2}$ is the minimal  coupling parameter for the 3 cases of G\"{o}del-type metric. This  means that the Dirac equation in  the Weyl representation is invariant under a gauge transformation \cite{shap} of this kind: $\psi' = e^{f(z)}\psi$ and $A'_{z} = A_{z} - \nu^{-1}\partial f(z)$. From the Dirac's phase method, we can write the solution for Dirac equation in the following way:
\begin{eqnarray}
\psi = exp\left[-i\nu\int^{z}_{z_0} A_z dz\right]\psi_0,
\label{3.1}
\end{eqnarray}  
the exponential term corresponds to relativistic phase acquired by wave function of the massless fermion, and $\psi_0$ is the solution of Dirac equation:
\begin{eqnarray}
i\sigma^0\frac{\partial\psi_0}{\partial t} &+& i\sigma^1\left(D(r)\frac{\partial}{\partial r} -\frac{D(r)}{J(r)}\frac{\omega^{\ 1}_{\phi\ 2}}{2} +\frac{H(r)}{J(r)}\frac{\omega^{\ 1}_{t\ 2}}{2}\right)\psi_0 +\nonumber\\
 &+& i\sigma^2\left(\frac{D(r)}{J(r)}\frac{\partial}{\partial \phi} -\frac{H(r)}{J(r)}\frac{\partial}{\partial t}\right)\psi_0 + i\sigma^3\frac{\partial\psi_0}{\partial z} = 0.
\label{3.2}
\end{eqnarray}
 Now we obtain the Hamiltonian of this system.
It's possible to rewrite the Weyl equation (\ref{3.2}) in the follow way:
\begin{eqnarray}
i\frac{\partial\psi_0}{\partial t} = \left[\sigma^{1}\hat{\pi}_{r} + \sigma^{2}\hat{\pi}_{\phi} + \sigma^{3}\hat{\pi}_{z}\right]\psi_0 = \hat{H}\psi_0,
\label{3.2.a}
\end{eqnarray}
where $H$ is the Hamiltonian of Weyl particle and  the conjugated momentum are given by,
\begin{eqnarray}
\hat{\pi}_{r} &=& -i\left(D(r)\frac{\partial}{\partial r} -\frac{D(r)}{J(r)}\frac{\omega^{\ 1}_{\phi\ 2}}{2} +\frac{H(r)}{J(r)}\frac{\omega^{\ 1}_{t\ 2}}{2}\right);\nonumber\\
\hat{\pi}_{\phi} &=& -i\left(\frac{D(r)}{J(r)}\frac{\partial}{\partial \phi} -\frac{H(r)}{J(r)}\frac{\partial}{\partial t}\right);\\
\hat{\pi}_{z} &=& -i\frac{\partial}{\partial z}.\nonumber
\end{eqnarray}

Analysing the symmetry of the Hamiltonian $H$ to solve the Weyl equation (\ref{3.2}) associate with $H$, we can choice  the following  {\it ansatz} :
\begin{eqnarray}
\psi_{0}\left(t,\,r,\,\phi,\,z\right)= \left(\begin{array}{c} \psi_1(r) \\ \psi_2(r)\end{array}\right)e^{-i\left(Et-\ell\,\phi-k\,z\right)}, 
\label{3.3}
\end{eqnarray}
where $\psi_1$ and $\psi_2$ are two-component spinors, and through the Pauli matrices: 
\begin{eqnarray}
\sigma^0 = \left(\begin{array}{cc} 1 & 0\\ 0 & 1\end{array}\right),\ \ \sigma^1 = \left(\begin{array}{cc} 0 & 1\\ 1 & 0\end{array}\right),\ \ \sigma^2 = \left(\begin{array}{cc} 0 & -i\\ i & 0\end{array}\right),\ \ \sigma^3 = \left(\begin{array}{cc} 1 & 0\\ 0 & -1\end{array}\right);
\label{3.5}
\end{eqnarray}

then we obtain two coupled differential equations.   In next section we solve the eigenvalues and eigenfunction associated to Hamiltonian $H$ for flat, spherical and hyperboluc G\"odel space-times witha topological defect.
\section{The solution of Weyl equation in the Som-Raychaudhuri geometry}\label{subsec1}
In this section we solve the Weyl equation for flat G\"odel metric.
Let  us start with the solution of Weyl equation in Som-Raychaudhuri spacetime. For this case, we have that $D(r) = 1$, $H(r)=\alpha\Omega r^{2}$ and $J(r) = \alpha r$. Therefore, the Weyl equation (\ref{3.2}) will assume the following form,
\begin{eqnarray}
i\sigma^0\frac{\partial\psi_0}{\partial t} + i\sigma^1\left(\frac{\partial}{\partial r} +\frac{1}{2r}\right)\psi_0 + \frac{i\sigma^2}{\alpha r}\left(\frac{\partial}{\partial \phi} -\alpha\Omega r^{2}\frac{\partial}{\partial t}\right)\psi_0 + i\sigma^3\frac{\partial\psi_0}{\partial z} = 0.
\label{3.A.1}
\end{eqnarray}

To solve the equation (\ref{3.A.1}), we must use the {\it ansatz} (\ref{3.3}), and the Pauli matrices (\ref{3.5}). Then we obtain two coupled differential equations: 
\begin{eqnarray}
\left[E - k\right]\psi_1 &= -i\left[\frac{\partial}{\partial r} + \frac{1}{2r} - \frac{\ell}{\alpha r} - E\Omega r\right]\psi_2\label{3.A.2};\\
\left[E + k\right]\psi_2 &= -i\left[\frac{\partial}{\partial r} + \frac{1}{2r} + \frac{\ell}{\alpha r} + E\Omega r\right]\psi_1.\label{3.A.3}
\end{eqnarray}

Now it is possible to decouple these two differential equations. From equations (\ref{3.A.2}) and (\ref{3.A.3}), we are capable to convert these two differential equations of first order to two  differential equations of second order. Thereby,
\begin{eqnarray}
\frac{d^2 \psi_i}{d r^2} + \frac{1}{r}\frac{d \psi_i}{d r} - \left[E^{2}\Omega^{2} r^{2} + \frac{\zeta^{2}_{i}}{r^2} - 4E\Omega\beta_i \right]\psi_i = 0,
\label{3.A.6}
\end{eqnarray}
with $i = 1, 2$ label each spinor component. And the  parameter $\zeta_i$ and $\beta_i$ is given by,
\begin{eqnarray}
\zeta_1 = \left(\frac{|\ell|}{\alpha} + \frac{1}{2}\right),\ \ \zeta_2 = \left(\frac{|\ell|}{\alpha} - \frac{1}{2}\right),
\label{3.A.7} 
\end{eqnarray}
and,
\begin{eqnarray}
\beta_1 &= \frac{E}{4\Omega} - \frac{k^{2}}{4E\Omega} + \frac{1}{4} - \frac{\ell}{2\alpha};\label{3.A.8}\\
\beta_2 &= \frac{E}{4\Omega} - \frac{k^{2}}{4E\Omega} - \frac{1}{4} - \frac{\ell}{2\alpha}.
\label{3.A.9} 
\end{eqnarray}

Let us do a change  of variables in  the equation (\ref{3.A.6}), where $\varsigma = E\Omega r²$. Thus,
\begin{eqnarray}
\varsigma\frac{d^{2}\psi_i}{d\varsigma^{2}} + \frac{d\psi_i}{d\varsigma} + \left[\beta_i - \frac{\varsigma}{4} - \frac{\zeta^{2}_{i}}{4\varsigma}\right]\psi_i = 0. 
\label{3.A.10}
\end{eqnarray}

The solution of equation (\ref{3.A.10}) is
\begin{eqnarray}
\psi_i(\varsigma) = e^{-\frac{\varsigma}{2}}\varsigma^{\frac{|\zeta_i|}{2}}F(\varsigma),
\label{3.A.11}
\end{eqnarray}
 substituting this solution in equation (\ref{3.A.10}), we  get
\begin{eqnarray}
\varsigma\frac{d^{2}F}{d\varsigma^{2}} +\left(|\zeta_i | + 1 - \varsigma\right) \frac{dF}{d\varsigma} + \left(\beta_i - \frac{1}{2} - \frac{|\zeta_{i}|}{2}\right)F = 0.
\label{3.A.12}
\end{eqnarray}

This  equation is the confluent hypergeometric differential equation, whose solution is the confluent hypergeometric function $F(\varsigma) = F\left(-\left[\beta_i - \frac{1}{2} - \frac{|\zeta_{i}|}{2}\right], |\zeta_i| + 1; \varsigma\right)$. Therefore, the first parameter of confluent hypergeometric function is
\begin{eqnarray}
\nu = \beta_i - \frac{1}{2} - \frac{|\zeta_{i}|}{2},
\label{3.A.13}
\end{eqnarray} 
so that the energy levels for both spinors  are given by,
\begin{eqnarray}
E_{\nu,\,\ell}&=&2\Omega\left[\nu+\frac{|\ell|}{2\alpha}+\frac{\ell}{2\alpha}+\frac{1}{2}\right]
\pm 2\Omega\sqrt{\left[\nu+\frac{|\ell|}{2\alpha}+\frac{\ell}{2\alpha}+\frac{1}{2}\right]^{2}+\frac{k^{2}}{4\Omega^{2}}},
\label{3.A.14}
\end{eqnarray} 
where $\nu=0,1,2,3,\ldots$ and $\ell$ is half-integer.   In this way, the expression (\ref{3.A.14}) gives the eigenvalues of energy of a Weyl fermion in the   Som-Raychaudhuri geometry pierced by  a topological defect. In the  limit $\alpha=1$, we obtain the same results for energy spectrum of Weyl fermions  arisen in the Som-Raychaudhuri metric,  note that we do not obtain  from the Eq. (\ref{3.A.14}) the spectrum found in the  Ref. \cite{gabriel}, due to the dependence of  torsion coupling obtained in energy spectrum of Ref. \cite{gabriel} even considering the limit case of  the massless fermions.  We conclude that the energy  spectrum  (\ref{3.A.14})  is similar  to the the relativistic Landau levels for fermions and  is characterized by an  infinite degeneracy for $\alpha=1$. Note that  the degeneracy of the eigenvalues (\ref{3.A.14}) is broken due  to the presence of topological defect ($\alpha\neq 1$) in this background. Now we analyze the zero mode energy for this case. Now we analyse the zero mode $E_{\nu=0}=0$ energy for this case.  From equation (\ref{3.A.14}) we can derive the zero modes  of the Weyl spinor in this G\"odel-type metric with the topological defect considering $\nu=0$. The zero mode occurs when the particle is confined to plane, i. e., when $k=0$.

Thus,  the eigenspinor  for the Weyl  particle in the Som-Raychaudhuri geometry with the topological defect is given by
\begin{eqnarray}
\psi_0\left(t,\,\varsigma,\,\phi,\,z\right)&=&C_{\nu,\ell }\,e^{-\frac{\varsigma}{2}}e^{-i\left[Et-\ell \phi-kz\right]}\times\nonumber\\
[-2mm]\label{3.A.15.a}\\[-2mm]
&\times&\left(\begin{array}{c}
\varsigma^{\frac{\ell}{\alpha} +\frac{1}{2}}\,_{1}F_{1}\left(-\nu,\,\frac{\ell}{\alpha}+\frac{3}{2};\,\varsigma\right)\\
\varsigma^{\frac{\ell}{\alpha}-\frac{1}{2}}\,_{1}F_{1}\left(-\nu,\,\frac{\nu}{\alpha}+\frac{1}{2};\,\varsigma\right)\\
 \end{array}\right),\nonumber
\end{eqnarray}
where $C_{\nu, \ell}$ is a constant spinor,  and $\ell \geqslant\frac{1}{2}$. The wave function for case  $\ell \leqslant -\frac{1}{2}$ is
\begin{eqnarray}
\psi_0\left(t,\,\varsigma,\,\phi,\,z\right)&=&C_{\nu, \ell}\,e^{-\frac{\varsigma}{2}}e^{-i\left[Et-\ell \phi-kz\right]}\times\nonumber\\
[-2mm]\label{3.A.15.b}\\[-2mm]
&\times&\left(\begin{array}{c}
\varsigma^{-\frac{\ell}{\alpha} -\frac{1}{2}}\,_{1}F_{1}\left(-\nu,\,-\frac{\ell}{\alpha}+\frac{1}{2};\,\varsigma\right)\\
\varsigma^{-\frac{\ell}{\alpha}+\frac{1}{2}}\,_{1}F_{1}\left(-\nu,\,-\frac{\ell}{\alpha}+\frac{3}{2};\,\varsigma\right)\\ \end{array}\right),\nonumber
\end{eqnarray}
here $C_{\nu, \ell}$ is a constant spinor.

\section{ Solution of Weyl equation in the spherically symmetric  G\"{o}del-type geometry  }\label{subsec2}
Now we investigate the Weyl equation  in the  spherically symmetric  G\"{o}del-type spacetime. 
 So, we will resolve the Weyl equation (\ref{3.2})  for $D(r) = \left(1 +\frac{r^{2}}{4R^{2}}\right)$, $H(r) = \alpha\Omega r^{2}$ and $J(r) = \alpha r$:
\begin{eqnarray}
i\sigma^{0}\frac{\partial\psi_0}{\partial t}&+&i\sigma^{1}\left[\left(1+\frac{r^{2}}{4R^{2}}\right)\frac{\partial}{\partial r} +\left(1-\frac{r^{2}}{4R^{2}}\right)\frac{1}{2r}\right]\psi_0 +\frac{i\sigma^{2}}{\alpha\,r}\left(1+\frac{r^{2}}{4R^{2}}\right)\frac{\partial\psi_0}{\partial \phi}\nonumber\\
[-2mm]\label{3.B.1}\\[-2mm]
&-&i\,\Omega\,r\,\sigma^{2}\,\frac{\partial\psi_0}{\partial t}+i\sigma^{3}\frac{\partial\psi_0}{\partial z} = 0,\nonumber
\end{eqnarray}

Using the spinor (\ref{3.3}) to solve the Dirac equation, we get two coupled equations: 
\begin{eqnarray}
\left[E - k\right]\psi_1 &= i\left[\left(1 + \frac{r^2}{4R^2}\right)\frac{\partial}{\partial r} +\left(1 - \frac{r^2}{4R^2}\right) \frac{1}{2r} - \left(1 + \frac{r^2}{4R^2}\right)\frac{\ell}{\alpha r} - E\Omega r\right]\psi_2;\label{3.B.2}\\
\left[E + k\right]\psi_2 &= i\left[\left(1 + \frac{r^2}{4R^2}\right)\frac{\partial}{\partial r} +\left(1 - \frac{r^2}{4R^2}\right)\frac{1}{2r} + \left(1 + \frac{r^2}{4R^2}\right)\frac{\ell}{\alpha r} + E\Omega r\right]\psi_1\label{3.B.3}.
\end{eqnarray}

It is possible to convert these equations, so that we obtain two second order equations for each spinor, so that we have
\begin{eqnarray}
\left(1 + \frac{r^2}{4R^2}\right)^2\left[\frac{d^2 \psi_i}{d r^2} + \frac{1}{r}\frac{d \psi_i}{d r}\right] - \left[a'_i r^2 + \frac{b^{2}_i}{r^2} - c_i \right]\psi_i = 0,
\label{3.B.6}
\end{eqnarray}
for $i = 1, 2$ representing each spinor. The parameters $a'_i$ are given by:
\begin{eqnarray}
a'_1 =& \frac{a^{2}_{1}}{16R^4} =\frac{1}{16R^4} \left(\frac{\ell}{\alpha} - \frac{1}{2} + 4R^{2}E\Omega  \right)^2;\label{3.B.7}\\ 
a'_2 =& \frac{a^{2}_{2}}{16R^4} =\frac{1}{16R^4} \left(\frac{\ell}{\alpha} + \frac{1}{2} + 4R^{2}E\Omega  \right)^2,\label{3.B.8}
\end{eqnarray}
and $b_i$ -- by
\begin{eqnarray}
b^{2}_{1} =& \left(\frac{\ell}{\alpha} + \frac{1}{2}\right)^2\label{3.B.9}\\
b^{2}_{2} =& \left(\frac{\ell}{\alpha} - \frac{1}{2}\right)^2.\label{3.B.10}
\end{eqnarray}
 Finally,
\begin{eqnarray}
c_1 &= E^2 - k^2 - \frac{3}{8R^2} + E\Omega - \frac{\ell^2}{2\alpha^2 R^2} - \frac{2\ell E\Omega}{\alpha}\label{3.B.11}\\
c_2 &= E^2 - k^2 - \frac{3}{8R^2} - E\Omega - \frac{\ell^2}{2\alpha^2 R^2} - \frac{2\ell E\Omega}{\alpha}.\label{3.B.12}
\end{eqnarray}

Now,  let us solve the differential equation (\ref{3.B.6}). We should introduce  the new coordinate $\theta$,  through a change of variables $r = 2R\tan\theta$. Thereby, the equation (\ref{3.B.6}) will be
\begin{eqnarray}
\frac{d^2 \psi_i}{d \theta^{2}} +\left( \frac{1}{\cos\theta\sin\theta} - \frac{2\sin\theta}{\cos\theta}\right)\frac{d\psi_i}{d\theta} - \left[a^{2}_{i}\frac{\sin^{2}\theta}{\cos^{2}\theta} + b^{2}_i \frac{\cos^{2}\theta}{\sin^{2}\theta} - 4R^{2}c_i \right]\psi_i = 0,
\label{3.B.13}
\end{eqnarray}
and we still can do two more changes of variables, first, $x = \cos\theta$, and  then, $\varsigma = 1 - x^2$,  hence we get
\begin{eqnarray}
\varsigma\left(1 -\varsigma\right)\frac{d^2 \psi_i}{d \varsigma^{2}} +\left( 1 - 2\varsigma\right)\frac{d \psi_i}{d\varsigma} - \left[\frac{a^{2}_{i}}{4}\frac{\varsigma}{\left(1 -\varsigma\right)} + \frac{b^{2}_i}{4} \frac{\left(1 -\varsigma\right)}{\varsigma} - R^{2}c_i \right]\psi_i = 0.
\label{3.B.14}
\end{eqnarray}

 Studying the asymptotic  limits of  (\ref{3.B.14}) we obtain  $\lambda_i=\frac{\left|a_{i}\right|}{2}$ and $\delta_i=\frac{\left|b_{i}\right|}{2}$, in this way, the solution  of  Eq. (\ref{3.B.14})  can be written  as
\begin{eqnarray}
\psi_{i}\left(\varsigma\right)=\varsigma^{\delta_i}\left(1-\varsigma\right)^{\lambda_i}\,\bar{F}_{i}\left(\varsigma\right), 
\label{3.B.15}
\end{eqnarray}
 where $\bar{F}_{i}\left(\varsigma\right)$  are unknown functions. By substituting the solution  (\ref{3.B.15}) into  (\ref{3.B.14}), we obtain the following equations for functions  $\bar{F}_{1}\left(\varsigma\right)$ and $\bar{F}_{2}\left(\varsigma \right)$,
\begin{eqnarray}
\varsigma\left(1-\varsigma\right)\frac{d^{2} \bar{F}_{i}}{d \varsigma^{2}}+\left[ 2\delta_i + 1 - 2\varsigma\left(\delta_i + \lambda_i + 1\right)\right]\frac{d \bar{F}_{i}}{d\varsigma}
 -\left[\lambda_{i} + 2\lambda_{i}\delta_{i} + \delta_{i} - R^{2}c_i\right] \bar{F}_{i}=0.
\label{3.B.16}
\end{eqnarray} 

 We have a set of two  decoupled  hypergeometric differential equations in  the Eqs. (\ref{3.B.16}), and the functions $\bar{F}_{i}\left(x\right)=\,_{2}F_{1}\left({\cal{A}},\,{\cal{B}},{\cal{ C}}_{i};\,\varsigma\right)$ are defined  with the coefficients $\cal{A}$ and $\cal{B}$ given by
\begin{eqnarray}
{\cal{A}}&=& \left[\left(\frac{\ell}{\alpha} + \frac{1}{2}\right) + 2R^{2}E\Omega\right] + \sqrt{4R^{4}\Omega^{2}E^{2} +  R^{2}\left[E^{2} - k^{2}\right]};\nonumber\\
\label{3.17}\\
\cal{B}&=& \left[\left(\frac{\ell}{\alpha} + \frac{1}{2}\right) + 2R^{2}E\Omega\right] - \sqrt{4R^{4}\Omega^{2}E^{2} + R^{2}\left[E^{2} - k^{2}\right]};\nonumber
\end{eqnarray}

Next, by imposing that the hypergeometric series  truncates, becoming a polynomial of  the degree $n$, then, we obtain (for both spinors)
\begin{eqnarray}\label{sphere} 
E_{\nu,\,\ell}&=& 2\Omega\left[\nu+\frac{|\ell |}{2\alpha}+\frac{\ell}{2\alpha} +\frac{1}{2}\right]\nonumber\\
&\pm& 2\Omega\left\{\left[\nu+\frac{|\ell |}{2\alpha}+ \frac{\ell }{2\alpha}+\frac{1}{2}\right]^{2}+\frac{1}{4R^{2}\,\Omega^{2}}\left[\nu+\frac{|\ell |}{2\alpha}+\frac{\ell }{2\alpha}+\frac{1}{2}\right]^{2} + \frac{k^{2}}{4\Omega^{2}}\right\}^{1/2},
\end{eqnarray}
where $\nu=0,1,2,3,\ldots$ and $-\nu +\frac{1}{2}\leqslant \frac{\ell }{\alpha}\leqslant 4\Omega R^2E + \frac{1}{2}$.  Note that, in this case we not observe a possibility of zero mode, for case where $R \longrightarrow \infty $, we obtain the flat case found in previous section for Som-Raychaudhuri metric. Note that the curvature and  the rotation  introduce a mass term in the Weyl fermion energy levels. Note that in the limit where $\alpha=1$ we obtain the eigenvalues for Weyl fermions in the background of  the spherical G\"odel metric. We can also observe that the results obtained in this section  differ from results obtained in the  zero mass limit in Ref. \cite{gabriel} for  Dirac fermions in a  spherical G\"odel spacetime with torsion.

The corresponding eigenspinor for  a Weyl particle in the spherical G\"odel-type geometry with a topological defect is given by
\begin{eqnarray}
\psi_0(t,\,\varsigma,\,\phi,\, z)&=&\bar{C}_{n, m}\,e^{-i\left[Et-\ell \phi-kz\right]}\times\\\nonumber
&\times&\left(\begin{array}{c}
\left(1-\varsigma\right)^{\frac{\left|a_{2}\right|}{2}}\,\varsigma^{\frac{\left|b_{2}\right|}{2}}\,_{2}F_{1}\left({\cal{A}},\,{\cal{B}},\,\frac{\ell }{\alpha}+\frac{3}{2};\,\varsigma\right)\\ 
\left(1-\varsigma\right)^{\frac{\left|a_{1}\right|}{2}}\,\varsigma^{\frac{\left|b_{1}\right|}{2}}\,_{2}F_{1}\left({\cal{A}},\,{\cal{B}},\,\frac{\ell }{\alpha} +\frac{1}{2};\,\varsigma\right)
\end{array}\right),
\label{3.B.20}
\end{eqnarray}
where $\bar{C}_{\nu, m}$ is a constant spinor  and the parameters ${\cal{A}}$ and ${\cal{B}}$ of the hypergeometric functions have been defined in Eq. (\ref{3.17}) and $\frac{1}{2}\leqslant \frac{\ell}{\alpha}\leqslant 4\Omega R^2E + \frac{1}{2}$. And we have
\begin{eqnarray}
\psi_0(t,\,\varsigma,\,\phi,\, z)&=&\bar{C}_{\nu, m}\,e^{-i\left[Et- \ell \phi-kz\right]}\times\\\nonumber
&\times&\left(\begin{array}{c}
\left(1-\varsigma\right)^{\frac{\left|a_{2}\right|}{2}}\,\varsigma^{\frac{\left|b_{2}\right|}{2}}\,_{2}F_{1}\left(1-{\cal{A}},\,1-{\cal{B}},\,-\frac{\ell}{\alpha}+\frac{1}{2};\,\varsigma\right)\\ 
\left(1-\varsigma\right)^{\frac{\left|a_{1}\right|}{2}}\,\varsigma^{\frac{\left|b_{1}\right|}{2}}\,_{2}F_{1}\left(1-{\cal{A}},\,1-{\cal{B}},\,-\frac{\ell}{\alpha} +\frac{3}{2};\,\varsigma\right)
\end{array}\right),
\end{eqnarray}
where $\bar{C}_{\nu, m}$ is a constant spinor and  $-\nu + \frac{1}{2}\leqslant \frac{\ell }{\alpha}\leqslant -\frac{1}{2}$.
\section{Solution of Weyl equation in the Hyperbolic G\"{o}del-type geometry}\label{subsec3}

Finally we will  analyse the third case  with use of the equation (\ref{3.2}) with $D(r) = \left(1 - l^2 r^2\right)$, $H(r) = \alpha\Omega r^{2}$ and $J(r) = \alpha r$,
\begin{eqnarray}
i\sigma^{0}\frac{\partial\psi_0}{\partial t}&+&i\sigma^{1}\left[\left(1-l^{2} r^{2}\right)\frac{\partial}{\partial r} +\left(1+l^{2} r^{2}\right)\frac{1}{2r}\right]\psi_0+\frac{i\sigma^{2}}{\alpha\,r}\left(1-l^{2} r^{2}\right)\frac{\partial\psi_0}{\partial \phi}\nonumber\\
[-2mm]\label{3.C.1}\\[-2mm]
&-&i\,\Omega\,r\,\sigma^{2}\,\frac{\partial\psi_0}{\partial t}+i\sigma^{3}\frac{\partial\psi_0}{\partial z} =0,\nonumber
\end{eqnarray}
and one more time we use the eigenstate given by (\ref{3.3}) to obtain two coupled differential equations:
\begin{eqnarray}
\left[E - k\right]\psi_1 &=& i\left[\left(1 - l^{2}r^{2}\right)\frac{\partial}{\partial r} +\left(1 + l^{2}r^{2}\right) \frac{1}{2r} - \left(1 - l^{2}r^{2}\right)\frac{\ell}{\alpha r} - E\Omega r\right]\psi_2;\label{3.C.2}\\
\left[E + k\right]\psi_2 &=& i\left[\left(1 - l^{2}r^{2}\right)\frac{\partial}{\partial r} +\left(1 + l^{2}r^{2}\right)\frac{1}{2r} + \left(1 - l^{2}r^{2}\right)\frac{\ell}{\alpha r} + E\Omega r\right]\psi_1\label{3.C.3}.
\end{eqnarray} 

We can convert these two coupled differential equation of first order in two decoupled differential equation of second order, as we made earlier. Therefore, we have two equations written in a compact way:
\begin{eqnarray}
\left(l^{2}r^{2} - 1\right)^2\left[\frac{d^2 \psi_i}{d r^{2}} + \frac{1}{r}\frac{d \psi_i}{d r}\right] - \left[a'_i r^{2} + \frac{b^{2}_i}{r^{2}} - c_i \right]\psi_i = 0,
\label{3.C.6}
\end{eqnarray}
each spinor is represented by $i = 1, 2$. And the parameters of equation (\ref{3.C.6}) are,
\begin{eqnarray}
a'_1 =& l^4 a^{2}_{1} = l^4\left(\frac{\ell}{\alpha} - \frac{1}{2} - \frac{E\Omega}{l^{2}}  \right)^2;\label{3.C.8}\\ 
a'_2 =& l^4a^{2}_{2} = l^4\left(\frac{\ell}{\alpha} + \frac{1}{2} - \frac{E\Omega}{l^{2}}\right)^2,\label{3.C.9}
\end{eqnarray}
and $b_i$,
\begin{eqnarray}
b^{2}_{1} =& \left(\frac{\ell}{\alpha} + \frac{1}{2}\right)^2\label{3.C.10}\\
b^{2}_{2} =& \left(\frac{\ell}{\alpha} - \frac{1}{2}\right)^2.\label{3.C.11}
\end{eqnarray}
Finally,
\begin{eqnarray}
c_1 = E^{2} - k^{2} + \frac{3l^{2}}{2} + E\Omega + \frac{2l^{2}\ell^{2}}{\alpha^{2}} - \frac{2\ell E\Omega}{\alpha}\label{3.C.12}\\
c_2 = E^{2} - k^{2} + \frac{3l^{2}}{2} - E\Omega + \frac{2l^{2}\ell^{2}}{\alpha^{2}} - \frac{2\ell E\Omega}{\alpha}.
\end{eqnarray}
 
To solve the equation (\ref{3.C.6}), we  make the change of the variable $r = \dfrac{\tanh(l\theta)}{l}$ and get
\begin{eqnarray}
\frac{d^2 \psi_{i}}{d \theta^{2}}&+&\left(\frac{2l\sinh\left(l\theta\right)}{\cosh\left(l\theta\right)}+ \frac{l}{\cosh\left(l\theta\right)\sinh\left(l\theta\right)}\right)\frac{d \psi_{i}}{d\theta}\nonumber\\
&-&\left[a^{2}_{i}l^{2}\frac{\sinh^{2}\left(l\theta\right)}{\cosh^{2}\left(l\theta\right)}+b^{2}_{i}\,l^{2}\,\frac{\cosh^{2}\left(l\theta\right)}{\sinh^{2}\left(l\theta\right)}-c_{i}\right]\psi_{i}=0,
\label{3.C.13}
\end{eqnarray}
  afterward, we make the following change of variables:   $y=\cosh(l\theta)$ and $\varsigma=y^{2}- 1$, in this way we obtain the following equation
\begin{eqnarray}
\varsigma\left(1+\varsigma\right)\frac{d^2 \psi_{i}}{d \varsigma^{2}} +\left(1+2\varsigma\right)\frac{d \psi_{i}}{d\varsigma}-\left[\frac{a^{2}_{i}}{4}\frac{\varsigma}{\left(1+\varsigma\right)}+\frac{b^{2}_{i}}{4}\,\frac{\left(1+\varsigma\right)}{\varsigma}-\frac{\varepsilon^{2}-1}{4}\right]\psi_{i}=0,
\label{3.C.14}
\end{eqnarray}
where  $\frac{\varepsilon^{2}-1}{4}= \frac{c_{i}}{4l^{2}}$. We  choose  $\lambda_i =\frac{\left|a_{i}\right|}{2}$ and $\delta_i =\frac{\left|b_{i}\right|}{2}$. Therefore, considering the critical poits of equation we can  write the solution of  Eq. (\ref{3.C.14})  in the form 
\begin{eqnarray}
\psi_{i}\left(\varsigma\right)=\varsigma^{\delta_i}\,\left(1+\varsigma\right)^{\lambda_i}\,\tilde{F}_{i}\left(\varsigma\right),
\label{3.C.15} 
\end{eqnarray}
where $\tilde{F}_{i}\left(\varsigma\right)$ is  an unknown functions. Thus, we obtain the following equation for $\tilde{F}_{i}\left(x\right)$:
\begin{eqnarray}
\varsigma\left(1 -\varsigma\right)\frac{d^{2} \tilde{F}_{i}}{d \varsigma^{2}}+\left[2\delta_i + 1 -2\varsigma\left(\delta_i + \lambda_i + 1\right)\right]\frac{d \tilde{F}_{i}}{d\varsigma} -\left[\lambda_i + 2\lambda_i\delta_i +\delta_i +\frac{c_i}{4l^{2}}\right]\tilde{F}_{i}=0.
\label{3.C.16}
\end{eqnarray}

It results in  two hypergeometric  differential equations, where the functions $\tilde{F}_{i}\left(\varsigma\right)=\,_{2}F_{1}\left({\cal{A}},\, {\cal{B}},\,{\cal{C}};\,\varsigma\right)$ and the parameters ${\cal{A}}$ and ${\cal{B}}$ given by 
\begin{eqnarray}
{\cal{A}}&=&\left[\left(\frac{\ell}{\alpha} + \frac{1}{2}\right) - \frac{E\Omega}{2l^{2}}\right]+ \sqrt{\frac{E^{2}\Omega^{2}}{4l^{4}} - \frac{1}{4l^{2}}\left[E^{2} - k^2\right]}\nonumber\\
\label{3.C.18}\\
{\cal{B}}&=& \left[\left(\frac{\ell}{\alpha} + \frac{1}{2}\right) - \frac{E\Omega}{2l^{2}}\right]- \sqrt{\frac{E^{2}\Omega^{2}}{4l^{4}} - \frac{1}{4l^{2}}\left[E^{2} - k^2\right]}\nonumber
\end{eqnarray}

As in the previous section, we impose that the hypergeometric series  must truncate, reducing to a polynomial of degree $\nu$, then, we obtain the following allowed energies for both spinors:
\begin{eqnarray}
E_{\nu,\,\ell}&=& 2\Omega\left[\nu+\frac{|\ell|}{2\alpha}+ \frac{j}{2\alpha}+\frac{1}{2}\right]\nonumber\\
&\pm& 2\Omega\left\{\left[\nu+\frac{|\ell|}{2\alpha}+\frac{\ell}{2\alpha}+\frac{1}{2}\right]^{2}-\frac{l^{2}}{\Omega^{2}}\left[n+ \frac{|\ell|}{2\alpha}+\frac{j}{2\alpha} +\frac{1}{2}\right]^{2} + \frac{k^2}{4\Omega^{2}}\right\}^{1/2},
\label{3.C.19} 
\end{eqnarray}
where $\nu=0,1,2,3,\ldots$ and $-\nu+\frac{1}{2}\leqslant\frac{\ell}{\alpha}\leqslant \infty$. Note that in the  $\alpha=1$ limit we obtain the eigenvalues for Weyl fermions in the background of hyperbolic  G\"odel-type metric. We can also observe that in present case the results obtained in this section is different of the obtained in the limit where mass is zero in Ref. \cite{gabriel}  for Dirac fermions in a  hyperbolic  G\"odel spacetime with torsion.
Now we  verify the existence of zero mode for this Weyl spinor in hyperbolic G\"odel-type geometry. When we impose the condition for (\ref{3.C.19}) produces a eigenvalues with null energy. For this case, we obtain the following condition for  zero modes  that are given by 
\begin{eqnarray}
k = \pm l;
\end{eqnarray} 
when $\ell \leq 0$, or
\begin{eqnarray}\label{zeromode}
k = \pm 2l\left(\frac{\ell}{\alpha} + \frac{1}{2}\right);
\end{eqnarray}
for $\ell \geq 0$.
In this case we observe that the condition (\ref{zeromode}) depends on the $\alpha$ parameter. In this way, the presence of topological defect modify the condition for existence of zero mode. In Refs \cite{figueiredo, fiol, josevi}  it was demonstrated that  eigenvalues of energy  in the  hyperbolic geometry  have two contributions: the first is discrete spectrum  contribution given  by Eq. (\ref{3.C.19}),  and the other continuous energy spectrum  contribution, whose lower bounds limit is obtained  by the relation $-\nu+\frac{1}{2}\leqslant\frac{\ell}{\alpha}\leqslant -\frac{1}{2}$ . In this form, in the hyperbolic G\"odel-type metric with  a topological defect, the eigenvalues of energy  have two parts, one is discrete and other is continuous, and they are determined  by the  parameter $\varepsilon$ through the relation 
\begin{eqnarray}
\varepsilon^{2}=\frac{\left(\Omega^{2}-l^{2}\right)}{4l^{4}}\,E^{2} + \frac{k^2}{4l^{2}};
\label{3.C.20}
\end{eqnarray}
in this way, we have discrete eigenvalues of energy for $\varepsilon^{2}\,\geq\,1$ and a continuous eigenvalues  for $\varepsilon^{2}\,<\,1$.

 Let us study the eigenvalues of energy (\ref{3.C.19})  considering the parameters $\Omega$ and $l$. Thus, we consider  three cases:

\begin{enumerate}
	\item The case   $\Omega^{2}\,>\,l^{2}$, we have of  Eq. (\ref{3.C.20}) a discrete  set of eigenvalues of  the following condition for energy  given by
\begin{eqnarray}
E^{2}&\geq&\frac{l^{2}}{\Omega^{2} - l^{2}}\left[4l^{2} - k^2\right].
\label{3.C.21}
\end{eqnarray}

Otherwise, the relativistic spectrum of energy is continuous.
	
	\item  The second case: considering   $\Omega^{2}=l^{2}$, we have from Eq. (\ref{3.C.20}) a discrete  set of eigenvalues of energy from the following condition
\begin{eqnarray}
k^2\geq 4l^{2}.
\label{3.C.23}
\end{eqnarray}
Otherwise, the relativistic eigenvalues  spectrum is continuous.

    \item  Finally, in the third case we consider the condition $\Omega^{2}\,<\,l^{2}$  and obtain  from Eq. (\ref{3.C.20}) a discrete set of eigenvalues  of energy from the condition  
\begin{eqnarray}
E^{2}&\leq&\frac{l^{2}}{l^{2} -\Omega^{2}}\left[k^2 - 4l^{2}\right].
\label{3.C.24}
\end{eqnarray}

Otherwise, we also have that the eigenvalues for Weyl particle  of energy is continuous.
\end{enumerate}
 Thus, the  spinor  for the Weyl particle in  the hyperbolic  case  is given by
\begin{eqnarray}
\psi(t,\,\varsigma,\,\phi,\,z)&=&\tilde{C}_{\nu,\,m}\,\,e^{-i\left[Et-\ell \phi-kz\right]}\times\nonumber\\
&\times&\left(\begin{array}{c}
\left(1+\varsigma\right)^{\frac{\left|a_{2}\right|}{2}}\,\varsigma^{\frac{\left|b_{2}\right|}{2}}\,_{2}F_{1}\left({\cal{A}},\,{\cal{B}},\,\frac{\ell}{\alpha}+\frac{3}{2};\,\varsigma\right)\\
\left(1+\varsigma\right)^{\frac{\left|a_{1}\right|}{2}}\,\varsigma^{\frac{\left|b_{1}\right|}{2}}\,_{2}F_{1}\left({\cal{A}},\,{\cal{B}},\,\frac{\ell}{\alpha}+\frac{1}{2};\,\varsigma\right)
\end{array}\right),
\label{3.C.26}
\end{eqnarray}
where $\tilde{C}_{\nu,\ \ell}$ is a constant spinor  and the parameters ${\cal{A}}$ and ${\cal{B}}$ of the hypergeometric functions have been defined in Eq. (\ref{3.C.18}), and $\frac{1}{2}\leqslant\frac{\ell}{\alpha}\leqslant\infty$. We have the spinor given by 
\begin{eqnarray}
\psi(t,\,\varsigma,\,\phi,\,z)&=&\tilde{C}_{\nu,\,m}\,\,e^{-i\left[Et-\ell \phi-kz\right]}\times\nonumber\\
&\times&\left(\begin{array}{c}
\left(1+\varsigma\right)^{\frac{\left|a_{2}\right|}{2}}\,\varsigma^{\frac{\left|b_{2}\right|}{2}}\,_{2}F_{1}\left(1-{\cal{A}},\,1-{\cal{B}},\,-\frac{\ell}{\alpha}+\frac{1}{2};\,\varsigma\right)\\
\left(1+\varsigma\right)^{\frac{\left|a_{1}\right|}{2}}\,\varsigma^{\frac{\left|b_{1}\right|}{2}}\,_{2}F_{1}\left(1-{\cal{A}},\,1-{\cal{B}},\,-\frac{\ell}{\alpha}+\frac{3}{2};\,\varsigma\right)
\end{array}\right),
\end{eqnarray}
with $\tilde{C}_{\nu,\ell}$ as constant spinor and $-\nu +\frac{1}{2}\leqslant\frac{\ell}{\alpha}\leqslant-\frac{1}{2}$. 

\section{Conclusions}\label{sec3}
 We have solved the Weyl equation  for a family of G\"odel-type geometries pieced by a topological defect.  We  obtained the   corresponding Weyl  equations in the Som-Raychaudhury, spherical and hyperbolic G\"odel  metrics containing a topological defect that passes  along $z$-axis, and solved them exactly. In the case of the Som-Raychaudhury metric with a topological defect, we have obtained the allowed energies for this relativistic quantum system and  have shown an analogy between the relativistic energy levels and the Landau levels,  with the rotation plays the role of the uniform magnetic field  along the $z$-direction. We have also seen that the presence of the topological defect breaks the degeneracy of the Weyl particle energy levels. We have discussed a possibility of energy zero mode and we obtain the condition for this possibility.

In the case of the spherically symmetric G\"odel-type metric with a cosmic string, we have also obtained the corresponding Weyl equation and solved it analytically. We have obtained  the eigenvalues of energies of this Weyl fermions  system. We observe that the presence of topological defect  breaks the degeneracy of system. For the spherical symmetry case we demonstrated the absence of possibility of zero mode, but in the limit of $R \longrightarrow \infty$  we obtain the condition of the Som-Raychaudhury case.

We  have also  obtained and solved analytically the Weyl equation in the hyperbolic G\"odel-type geometry with a topological defect. We have shown  that the eigenvalues of  energies for the system can be  both discrete and  continuous. Moreover, the  eigenvalues  are similar  to the Landau levels in a hyperbolic space, where we observe that  the presence of the topological defect  breaks  the degeneracy of the  eigenvalues  of energy.  We also investigate the possibility of  a zero mode and we obtain condition for existence of these zero modes. 

The results obtained in this contribution for quantum dynamics of Weyl fermions in family of G\"odel-type metric in  general relativity  are similar to Landau levels for Weyl fermions in a curved space. The energy levels obtained  for the class of  spacetimes has properties  different from the energy levels obtained in the case of quantum dynamics of Dirac fermions in the presence of G\"odel-type background in  theory relativity with torsion \cite{gabriel}. 

Note that the rotation of G\"odel spacetime  introduce  a contribution in the Weyl energy levels, for all case analysed here, similar  to a mass term.  This geometric mass term reduces a possibility of  an energy zero mode. We have  obtained  in the flat and hyperbolic case  a condition of existence of the zero mode  for Weyl particle in G\"odel-type metric  The present study can be important  to investigate the Hall effect for Weyl fermions in the  three-sphere $S^3$ \cite{nair,nair2} with presence of topological defect and  the higher dimensional quantum Hall effects and  to study the A-class topological insulators with emphasis on the noncommutative geometry \cite{hasebe}.  We  claim the analogy between energy levels for Weyl fermions in  a family of G\"odel-type metrics and Landau levels in curved spaces can be used to investigate the  Weyl semimetals  \cite{bin} in curved spaces. The systems investigated here can be used to describe condensed matter systems in curved geometries on the influence of rotation described by massless fermions\cite{everton,everton1,jonas,jonas1,brandao}

\acknowledgments{The authors would like to thank the Brazilian agencies CNPq, CAPES and FAPESQ for financial support.}



\begin{thebibliography}{99}
\addcontentsline{toc}{chapter}{Referências Bibliográficas}

\thispagestyle{myheadings}

\bibitem{godel} K. G\"odel, {\it Rev. Mod. Phys.} {\bf 21}, 447 (1949).
\bibitem{hawking} S. Hawking, {\it Phys. Rev. D} {\bf 46}, 603 (1992).
\bibitem{reboucas} M. Rebou\c{c}as, J. Tiomno, {\it Phys. Rev. D} {\bf 28}, 1251 (1983). 
\bibitem{reboucas2} M. Rebou\c{c}as, M. Aman, A. F. F. Teixeira, {\it J. Math. Phys.} {\bf 27}, 1370 (1985).
\bibitem{reboucas3} M. O. Galvao, M. Rebou\c{c}as, A. F. F. Teixeira, W. M. Silva, Jr, {\it J. Math. Phys.} {\bf 29}, 1127 (1988).
\bibitem{dabrowski} M. Dabrowski, J. Garecki, {\it Class. Quant. Grav.} {\bf 19}, 1 (2002), gr-qc/0102092.
\bibitem{dan}D Israel {\it JHEP} 0401, 042  (2004), hep-th/0310158

\bibitem{bertolami}O. Bertolami, F. Lobo, {\it Neuro. Quantol.} {\bf 7}, 1 (2009), arXiv: 0902.0559 [gr-qc].

\bibitem{barrow} J. Barrow, M. Dabrowski, {\it Phys. Rev. D} {\bf 58}, 103502 (1998), gr-qc/9803048; P. Kanti, C. E. Vayonakis, 
{\it Phys. Rev. D} {\bf 60}, 103519 (1999), gr-qc/9905032.

\bibitem {barrow2}J. D. Barrow, C. Tsagas, {\it Class. Quant. Grav.} {\bf 21}, 1773 (2004), gr-qc/0308067; 
{\it Phys. Rev. D} {\bf 69}, 064007 (2004), gr-qc/0309030; T. Clifton, J. Barrow, {\it Phys. Rev. D} {\bf 72}, 123003 (2005), gr-qc/0511076.
\bibitem{joeljmp}J. B. Fonseca-Neto, M. J. Rebou\c{c}as and A. F. F. Teixeira, {\it J. Math. Phys.} {\bf  33}, 2574
(1992).
\bibitem{joelrebou} J. B. Fonseca-Neto and M. J. Rebou\c{c}as, {\it Relativity and Gravitation,} {\bf 30}, 1301 (1998).
\bibitem{joel}  J. E. Aman, J. B.  Fonseca-Neto, M. A. H.  MacCallum, and M. J. Rebou\c{c}as,  {\it Class. Quant. Grav.} {\bf 15}, 1089 (1998). 
\bibitem{reboupala}J. Santos, M. J. Rebou\c{c}as, T. B. R. F. Oliveira, A. F. F. Teixeira, arXiv: 1611.03985.
\bibitem{furgodel}C. Furtado {\it et al.}, {\it Phys. Rev. D} {\bf 79}, 124039 (2009).
\bibitem{furgodel3}C. Furtado {\it et al.},  {\it Int. J. Mod. Phys. Conf. Ser.} {\bf 18},  145 (2012).
\bibitem{furgodel2}C. Furtado, J. R. Nascimento, A. Yu. Petrov and A. F. Santos, {\it Phys. Rev. D} {\bf 84},  047702 (2011).

\bibitem{joelhorava}J. B. Fonseca-Neto, A. Yu. Petrov and M. J. Rebou\c{c}as, {\it Phys. Lett. B} {\bf 725},  412 (2013).
\bibitem{porfi}J. A. Agudelo {\it et al.}, {\it Phys. Lett. B} {\bf  762}, 96 (2016).
\bibitem{everton}E. Cavalcante, J. Carvalho and  C. Furtado, {\it Eur. Phys. J. Plus} {\bf  131}, 288 (2016).


\bibitem{figueiredo} B. D. B. Figueiredo, I. D. Soares, J. Tiomno, {\it Class. Quant. Grav}. {\bf9}, 1593 (1992). 
 \bibitem{comtet} A. Comtet, {\it Ann. Phys.} {\bf 173}, 185 (1987). 
\bibitem{dunne} G. V. Dunne, {\it Ann. Phys.} {\bf 215}, 233 (1992).

\bibitem{fiol} N. Drukker, B. Fiol and J. Sim\'on, {\it JCAP} {\bf 0410}, 012 (2004).

\bibitem{bfiol} N. Drukker, B. Fiol and J. Sim\'{o}n, {\it Phys. Rev. Lett.} {\bf 91}, 231601 (2003).

\bibitem{gegenberg} S. Das and J. Gegenberg, {\it Gen. Rel. Grav.} {\bf 40}, 2115 (2008).
\bibitem{josevi} J. Carvalho, A. M. de M. Carvalho, C. Furtado, The Euro. Phys. Jour. C {\bf74}, 2935 (2014). 
\bibitem{epjpchina} Zhi Wang, Zheng-wen Long, Chao-yun Long, Ming-li Wu, Eur. Phys. J. Plus {\bf 130} 36 (2015).
\bibitem{vilalba} V.  M. Villalba, {\it Mod. Phys. Lett. A} {\bf 08}, 3011 (1993).
\bibitem{pimentel} L. O. Pimentel, A.  Camacho, and A. Macias, {\it Mod. Phys. Lett. A} {\bf  09}, 3703 (1994).
\bibitem{sandro}  S. G. Fernandes, G. de A. Marques and V. B. Bezerra,  {\it Class. Quant. Grav.} {\bf 23}, 7063 (2006).
\bibitem{havare}  A. Havare and T. Yetkin,  {\it Class. Quant Grav.} {\bf 19}, 1 (2002).
\bibitem{cohem} J. M.  Cohen,  C. V. Vishveshwara and  S. V. Dhurandhar  {\it J. Phys. A} {\bf 13}, 933 (1980).
\bibitem{hisco} W. A.  Hiscock, {\it Phys. Rev. D} {\bf 17}, 1497 (1978).
\bibitem{mash} B.  Mashoon,  {\it Phys. Rev. D} {\bf 11}, 2679 (1975).
\bibitem{pimentel1} L. O.  Pimentel  and A.  Macias, {\it Phys. Lett. A} {\bf 117}, 325 (1986). 
\bibitem{gabriel} G. Q. Garcia, J. R. de Oliveira, K. Bakke and C. Furtado, {\it Eur. Phys J. Plus} {\bf 132}, 123 (2017).
\bibitem{halp} B. I. Halperin, {\it Phys. Rev. B} {\bf 25}, 2185 (1982).

\bibitem{naka} M. Nakahara, {\it Geometry, Topology and Physics} (Cambridge University Press, Cambridge, UK, 1982).

\bibitem{weinberg} S. Weinberg, {\it Gravitation and Cosmology: Principles and Applications of the General Theory of Relativity} (IE-Wiley, New York, 1972).

\bibitem{cartan1} E. Cartan, {\it The Theory of Spinors} (Dover Publications, New York, 1981).


\bibitem{cartan2} E. Cartan, {\it Riemannian Geometry in an Orthogonal Frame} (World Scientific Publishing, Singapore, 2001).

\bibitem{carroll} S. Carroll, {\it Spacetime and Geometry; An Introduction to General Relativity} (Addison-Wesley Publishing Company, San Francisco, 2003).

\bibitem{shap} I. L. Shapiro, {\it Phys. Rep.} {\bf 357}, 113 (2002); L. H. Ryder {\bf and} I. L. Shapiro, {\it Phys. Lett. A} {\bf247}, 21 (1998).



\bibitem{nair} D. Karabali and  V. P. Nair, {\it Nucl. Phys.}  {\bf B641},  533 (2002), hep-th/0203264.
\bibitem{nair2} D. Karabali and  V. P. Nair, {\it Nucl. Phys.} {\bf  B679}, 427 (2004), hep-th/0307281.
\bibitem{hasebe}K. Hasebe, {\it Nucl. Phys.} {\bf B886}, 952 (2014).


\bibitem{bin}B. Yan and C. Felser, {\it Annu. Rev. Condens. Matter Phys.} {\bf 8}, 1 (2017). 

\bibitem{pal} P. Pal, {\it Amer. Jour. Phys.} {\bf 79}, 485 (2011).
\bibitem{everton1} G. Q Garcia, E. Cavalcante, A. M. de M. Carvalho and C. Furtado,  Eur. Phys. J. Plus  132, 183 (2017)
\bibitem{jonas} J. R. F. Lima, J. Brand\~ao, M. M. Cunha, F. Moraes,  Eur.  Phys J.  D {\bf 68}, 94 (2014).
\bibitem{jonas1} J. R. F. Lima and F. Moraes,  Eur.  Phys J.  B {\bf 88}, 263 (2015).
\bibitem{brandao} M. M. Cunha,  J. Brand\~ao, J. R. F. Lima and F. Moraes, Eur.  Phys J.  B {\bf 88}, 288 (2015).
\end{thebibliography}
\end{document}